\documentstyle[prl,aps,multicol,epsfig]{revtex}

\begin{document}
 
\title{Quasinormal Modes of Dirty Black Holes}
 
\author{P.T. Leung${}^{(1)}$, Y. T. Liu${}^{(1)}$,
W.-M. Suen${}^{(1, 2)}$,  C. Y. Tam${}^{(1)}$  and K. Young${}^{(1)}$}
 
\address{${}^{(1)}$Department of Physics, 
The Chinese University of Hong Kong, Hong Kong}
\address{${}^{(2)}$McDonnell Center for the Space Sciences, Department of Physics, Washington University, 
St Louis, MO 63130, U S A}

\date{\today}
 
\maketitle

\begin{abstract}
 
Quasinormal mode (QNM) gravitational radiation from black holes is
expected to be observed in a few years.  A perturbative formula is
derived for the shifts in both the real and the imaginary part of the
QNM frequencies away from those of an idealized isolated black hole.  The
formulation provides a tool for understanding how the astrophysical
environment surrounding a black hole, e.g., a massive accretion disk,
affects the QNM spectrum of gravitational waves.  We show, in a simple
model, that the perturbed QNM spectrum can have interesting features.

\end{abstract}
 
\draft

 
\begin{multicols}{2}

{\it 1. Introduction.}
The new generation of gravitational wave observatories (LIGO, VIRGO)
will soon be able to probe 
black holes in their dynamical interactions with the astrophysical
environment (e.g., matter or another black hole falling into it).
Numerical simulations \cite{num} show that the gravitational waves emitted in this process will
carry a signature associated with
the well-defined quasinormal mode (QNM) frequencies
of the black hole, and thereby
confirm its existence.

A stationary neutral black hole in an
otherwise empty and asymptotically flat spacetime
is a Kerr
hole (Schwarzschild hole in the case of zero angular momentum) \cite{nohair}.
Linearized gravitational waves propagating on the Kerr or Schwarzschild
background can be described by the Klein-Gordon equation \cite{chand}:

\begin{equation}
\left[ {\partial_{t}^{2}} - {\partial_{x}^{2}} +V(x) \right] \phi
(x,t) = 0 ,
\label{eq:kg}
\end{equation}

\noindent
where $x$ is a radial (tortoise) coordinate, $\phi$ is the radial part of a
combination of the metric functions representing the gravitational
wave.  The potential $V(x)$ describes the
scattering of the gravitational waves by the background geometry.
The outgoing wave boundary condition is appropriate for
waves escaping to infinity, and a
monochromatic solution [$\phi \propto \exp(-i \omega t)$]
is a QNM, with Im $\omega < 0$.  The QNM
spectra of Kerr and Schwarzschild black holes have been extensively
studied \cite{chand}, and provide a
template against which one can try to determine the nature of the
source; for an isolated black hole, the no-hair theorem \cite{nohair}
implies that the QNM spectrum depends only on the
mass $M$ and the angular momentum $J$.

However, the black holes that are observed will not be isolated, but will
be situated at the centers of galaxies, or will be surrounded by
accretion disks.  Therefore the observed spectra should not be
matched against those of a pure Kerr or Schwarzschild hole, but
to a black hole perturbed by interactions with its surrounding --- a
dirty black hole.  So far, the perturbation of black hole QNMs has
attracted little attention, partly because a perturbative formalism
for the QNMs of open systems, as opposed to the normal modes (NMs)
of conservative systems, has not hitherto been available.  In this
paper we develop such a formalism, which then opens the way to
inferring the astrophysical environment of the
black holes from the observed signal, beyond $M$ and $J$.

Two kinds of perturbations are involved here.  In the standard black
hole perturbation theory \cite{chand}, (\ref{eq:kg}) is obtained by
linearizing the metric about the Kerr or Schwarzschild background, and
the time-independent eigenvalue problem (with the outgoing wave
boundary condition) determines the QNM spectrum.  The
second type of perturbations are {\it the perturbations that change
the background} on which the wave propagates, e.g., by the presence of
an accretion disk; these  are often quasi-static, and hence separable
from that of the gravitational wave perturbation by the time scales
involved (in a suitable gauge choice).   In
this paper we focus on time-independent perturbation of the
background, described by (\ref{eq:kg}) with a potential
$V(x) = V_0(x) + \mu V_1(x)$, $|\mu| \ll 1$.  Therefore we are
led to study the following
eigenvalue problem in powers of $\mu$:

\begin{equation}
- \phi''(x) + \left[ V_0(x) + \mu V_1(x) \right] \phi = \omega^2 \phi 
\label{eq:kg1}
\end{equation}

While reminiscent of standard textbook problems, e.g., the usual
Rayleigh-Schr\"{o}dinger perturbation theory (RSPT), the
problem here is fundamentally different: the outgoing wave
condition renders the system physically nonconservative (energy
escapes to infinity) and the associated operator $ - d^2/dx^2 +V(x)$ 
non-hermitian; hermiticity underpins the usual RSPT.

The difficulty can be seen in several guises if one tries naively to
transcribe the usual formulas.  The first-order shift cannot be given by the
usual formula $\langle \phi_0 | \mu V_1 | \phi_0 \rangle / \langle
\phi_0 | \phi_0 \rangle$, in obvious notation --- the usual inner
product leads to $\langle \phi_0 | \phi_0 \rangle = \mbox{\Large
$\int$}_{-\infty}^{\infty} dx \phi_0^* \phi_0 = \infty$ since a QNM $
\phi_0$ extends over all space (and indeed grows exponentially at
infinity).  Higher-order shifts are even more problematic, since
the usual RSPT formula involves a sum over intermediate eigenstates,
but now the unperturbed eigenstates do not in general form a complete basis
\cite{kg-comp}, at least not in the case of black holes.  

{\it 2. Formulation.}  Our formulation
generalizes the logarithmic perturbation theory (LPT) \cite{3d} to
QNM systems; LPT has the property that it does not
require a complete set of eigenstates. 
Attention is focussed on the logarithmic derivative
$f(x)=\phi'(x) / \phi(x)$.  From (\ref{eq:kg1})

\begin{equation}
f'(x) + f^2(x) - \left[ V_0(x) + \mu V_1(x) \right] + \omega^2 = 0
\label{eq:ricc}
\end{equation}

\noindent
For any $\omega$, we define two solutions $f_{\pm}(\omega,x)$ by
the boundary conditions $f_{\pm}(\omega,x) \rightarrow \pm i\omega$
as $x \rightarrow \pm \infty$.  At an eigenvalue $\omega$, 
$f_+(\omega,x) = f_-(\omega,x)$.

For many cases of interest, $V(x) = V_0(x) + \mu V_1(x)$ 
is nontrivial only in a finite domain $(L_-,L_+)$, and is
relatively simple in the asymptotic
regions $(-\infty, L_-)$ and $(L_+,\infty)$.
In particular, we assume that the asymptotic regions can be solved
with the outgoing wave conditions to give the logarithmic derivatives
$D_{\pm}(\omega) = f_{\pm}(\omega,L_{\pm})$.  We then expand all
quantities
in powers of $\mu$: $ f \equiv f_0 + g = f_0 + \mu
g_1 + \mu^2 g_2 + \cdots$; $\omega = \omega_0 + \mu
\omega_1 + \cdots$; $D_{\pm} = D_{\pm 0} + \mu D_{\pm 1}+ \cdots$.

While the details of the derivation will be given elsewhere, the
central result for the $n$th order shift is 

\begin{equation}
\omega_n = \frac{ \langle \phi_0 | V_n | \phi_0 \rangle} 
{ 2 \omega_0 \langle \phi_0 |  \phi_0 \rangle } 
\label{eq:result}
\end{equation}

\noindent
in which we have introduced the suggestive notation 

\begin{eqnarray}
\langle \phi_0 | V_n | \phi_0 \rangle  &= &
\int_{L_-}^{L_+} \mbox{\Large{}}V_n(x) \phi_0^2(x) dx \nonumber \\
&-& \Delta_{+n} \phi_0^2(L_+) + \Delta_{-n} \phi_0^2(L_-)
\label{eq:matelm}
\end{eqnarray}

\begin{eqnarray}
\langle \phi_0 | \phi_0 \rangle &=&
\int_{L_-}^{L_+} \phi_0^2(x) dx \nonumber \\
&+& \frac{1}{2\omega_0} \left\{ D_{+0}' \phi_0^2(L_+)
- D_{-0}' \phi_0^2(L_-) \right\}
\label{eq:norm}
\end{eqnarray}

\noindent 
Here $V_1$ is the perturbing potential in (\ref{eq:kg1}), and for $n>1$, 
$V_n(x)=-\sum_{i=1}^{n-1} \left[ g_i(x) g_{n-i}(x) + \omega_i \omega_{n-i}
\right]$, with 

\begin{eqnarray}
\phi_0^2(x) g_n(x) =  & & \left[
\omega_n D_{-0}'(\omega_0) + \Delta_{-n} \right] \phi_0^2(L_-)
\nonumber \\
&+&
\int_{L_-}^x dy \left[ V_n(y) - 2\omega_0 \omega_n
\right] \phi_0^2(y)
\label{eq:gn}
\end{eqnarray}  

\noindent
Here $\Delta_{\pm n}$ is the
$n$th-order part of $D_{\pm}(\omega)-D_{\pm 0}(\omega_0)$; explicitly
$\Delta_{\pm 1} = D_{\pm 1}$, $\Delta_{\pm 2} = D_{\pm 2}+
\omega_1 D_{\pm 1}' + \frac{1}{2} \omega_1^2 D_{\pm 0}'' $  etc.,
where all $D_{\pm n}$ and their derivatives are understood to be
evaluated at the unperturbed frequency $\omega_0$.  These results
express the $n$th order correction to the eigenvalue in quadrature in
terms of lower-order quantities.  One can hence in principle obtain the
corrections to any order.  Similar to LPT for conservative systems, a
sum over intermediate states is not needed.

{\it 3. Properties of the Perturbed Spectrum For Open Systems in General.}
The result in (\ref{eq:result}) has been written in a way
formally similar to the hermitian case.
The factor $2\omega_0$
occurs because the eigenvalue is $\omega^2$ rather than $\omega$.
The numerator and the denominator in (\ref{eq:result}) are
separately independent of  $L_{\pm}$, so that they can be given
physical interpretations as a
generalized matrix element and
a generalized norm respectively.

The generalized norm
has some unusual properties \cite{waveeq,two-comp}.  
(a) It involves $\phi_0^2$ rather than $|\phi_0|^2$, and is in general complex.  
(b) It involves surface terms at $x = L_{\pm}$, though the value of the entire expression
is independent of the choice of $L_{\pm}$.  Thus, it is not a 
norm in the strict sense, but rather a useful bilinear map.
Nevertheless,
in cases where the system parameters can be tuned so that the
leakage of the wavefunction approaches zero (e.g., $V_0(x)$
contains tall barriers on both sides), 
the generalized norm does reduce to the usual (real and positive-definite)
norm for a NM.  

It is useful to define a function $H(x)$ for each QNM which depends only on the
original unperturbed system

\begin{equation}
\frac{\delta \omega}{\delta ( \mu V_1(x))} \equiv H(x) 
= \frac{\phi_0(x)^2}{ 2 \omega_0 \langle \phi_0 | \phi_0 \rangle }
\end{equation}

\noindent
Both the magnitude and the phase of $H(x)$ are well defined and
physically significant.  The magnitude implies that we can now give a
precise meaning to the normalization of a QNM, even though the
wavefunction diverges at infinity.  The phase of $H$ determines the
phase of the first-order shift $\omega_1$ for a real and positive
localized perturbation $V_1(x)$.  The phase is intriguing because it
has no counterpart for a hermitian system --- in that case, $H(x)$
must be real and non-negative.  

The functions $H(x)$ are then convenient objects for discussing the effect of
any perturbation on the QNMs of a given system.  We next present some properties
of $H(x)$ for the Schwarzschild black hole.

{\it 4. General Properties of the Perturbed Spectrum of a Schwarzschild Hole.}
Waves propagating on the exact Schwarzschild background geometry is 
described by (1) with the potential \cite{chand}

\begin{equation}
V_{Sc} (M,x) = \left( 1 - \frac{2M}{r} \right)
\left[ \frac{l(l+1)}{r^2} + (1 - s^2) \frac{2M}{r^3} \right]
\label{schw}
\end{equation}

\noindent
with $x = r + 2M \ln \left( r/2M -1 \right)$, where 
$s$ is the spin of the field ($s=2$ for
gravitational waves). 

In Fig. 1 we plot for the $s=0, l=1$ case the functions $H(x)$, which
{\it depends only on the unperturbed potential}.  The diagrams refer to the
lowest QNMs (labeled as $j = 0, 1, \cdots , 5$).  We note that for a
localized perturbation $\mu V_1(x) = \mu \delta(x-x_1)$, the frequency 
shift $\omega_1$ is given by $H(x_1)$, and therefore can be read
out directly from the figure.
Both $\mbox{Re $H(x)$}$ and $\mbox{Im $H(x)$}$
alternate in sign as $(-1)^j$ near the event horizon.  The patterns
are different for different values of $x_1$ and not simple,
demonstrating that a localized perturbation will push the QNMs along
different directions in the complex frequency plane,
generating a rich pattern of frequency shifts
(in contrast to shifts all of the same phase in the case of 
the NMs of a conservative system).  This 
implies much better prospects for extracting information
about the perturbing potential from the observed shifts.

\begin{figure}
\centerline{\epsfig{file=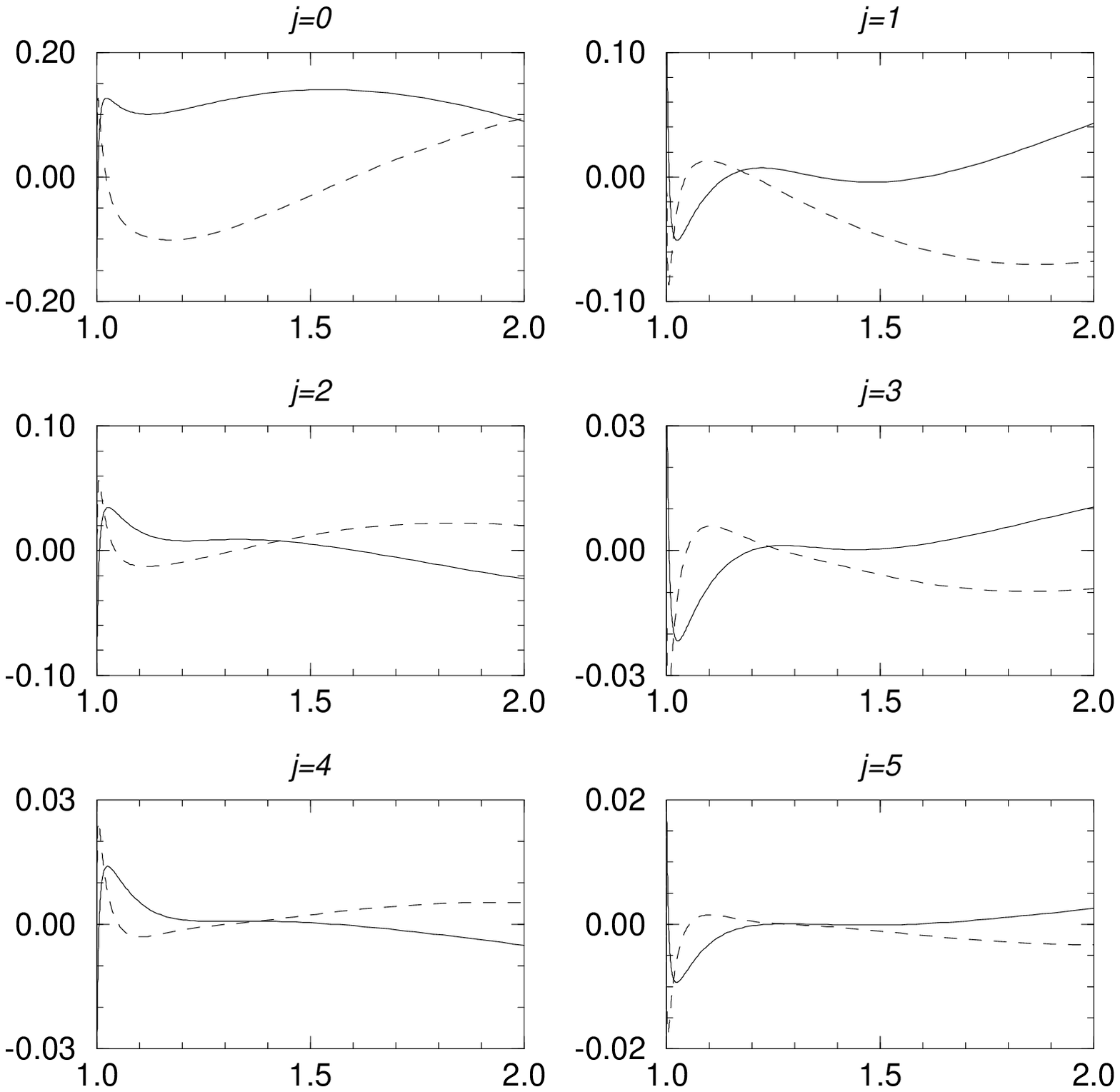,height=7cm}}
\end{figure}

\vskip -0.5cm
\noindent
{\small Fig. 1 \ \ Graph of Re $\left[ H(x)e^{-2\gamma \sqrt{1+(x/2M)^2}} 
\right]$
(solid line) and 
Im $\left[ H(x)e^{-2\gamma \sqrt{1+(x/2M)^2}} \right]$ (dashed line) vs
$r/2M$, where $\gamma$=Im($-2M\omega$)}

\vskip 0.5cm
The richness of the pattern could be diluted
if the perturbation has a spatial
extent $\Delta x$ large compared
to the typical wavelength of oscillation of 
$H(x)$, $\lambda \approx 2\pi / |\mbox{Re $\omega_0$}| 
\approx $ a few $M$.  Next we
discuss a model problem with an effective potential which
extends over an infinite range of $x$.

{\it 5. Perturbed Spectrum of a Schwarzschild Black Hole in a Model Problem.}
Consider a Schwarzschild hole surrounded by a static shell of matter.
Denote the total mass of the system as measured at infinity (ADM mass)
by $M_o$, and the
mass of the black hole as measured by its horizon surface area by $M_a$.
The perturbation is characterized by
$\mu \equiv (M_o-M_a)/M_a$ and the circumferential radius $r=r_s$
where the shell is placed.
For scalar wave ($s=0$), both
the unperturbed potential $V_0$ and the perturbation $V_1$ can be given in
terms of $V_{Sc}$ in (\ref{schw}): $V_0 (x)=V_{Sc}(M_o, x)$;
$ \mu V_1 (x) = \kappa \delta (x - x_s) + (\beta / \alpha) V_{Sc}(M_a, x) -
V_{Sc}(M_o, x)$,
for $x < x_s$, and $V_1=0$ for  $x > x_s$, where $x_s$ is the tortoise
coordinate at $r_s$.
The constants 
$\kappa , \alpha$ and $\beta$ are given by $M_o, \mu $ and $r_s$ in
some complicated expressions.
This perturbation consists of a $\delta$-function at 
the shell, plus a contribution inside the shell extending all the way
to the horizon ($x \rightarrow -\infty$,  
$r \rightarrow 2M_a$). 
There is no perturbation
outside the shell; in terms of the ADM mass,
the outside metric is exactly that of a Schwarzschild hole with $M_o$.

For $x<0$, the full potential $V=V_0 + \mu V_1$ can be expressed as 
a sum of exponentials, for which (\ref{eq:kg1}) with the
outgoing wave boundary
condition can be integrated analytically, thus giving the
log derivatives $D_-$, whereas the log derivative
$D_+$ is trivial because the perturbation vanishes outside the shell.  The details of the treatment of 
exponential potentials will be given elsewhere \cite{expon}. 

We first demonstrate the convergence of the perturbation results.
Fig. 2 shows the magnitude of the error in the frequencies in the 0th,
1st and 2nd order results versus $\mu$, for $l=1$ scalar waves,
compared to the exact numerical results (which 
can be obtained by brute force in this
simple case \cite{others}.)  The error of the $n$th order result
goes as $\mu^{n+1}$, as it should.  Detailed estimation of the error
and the large $n$ behavior of the perturbation expansion will be given
elsewhere.

\begin{figure}
\centerline{\epsfig{file=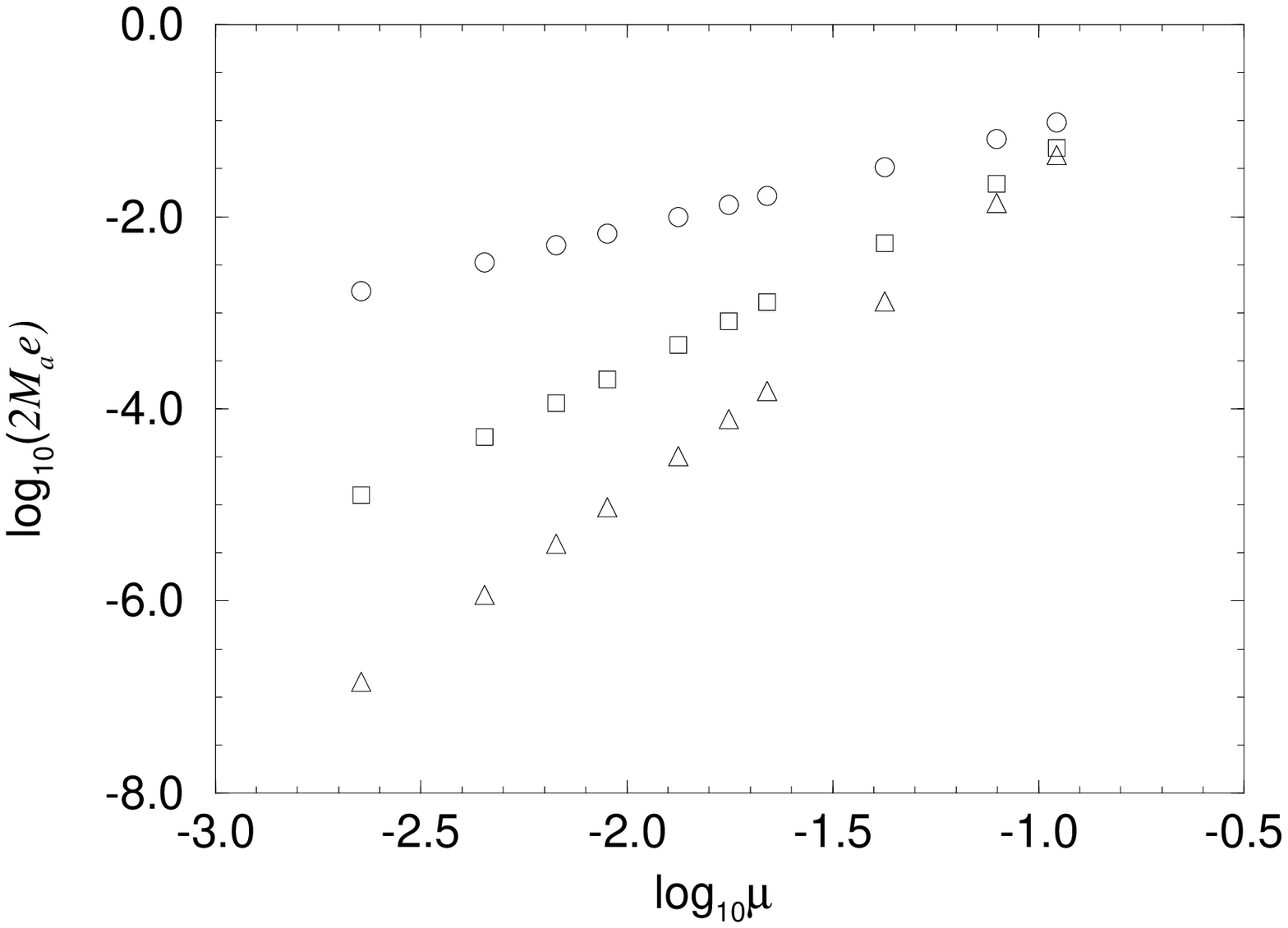,height=5cm}}
\end{figure}

\vskip -0.5cm
\noindent
{\small 
Fig.  2 \ \ The magnitude of the error $e$ in the frequencies of
the 0th (circles), 1st (squares) and 2nd order (triangles) perturbation for 
$l=1$, $s=0$, $j=1$, and $r_s=2.52 M_a$, vs the size of the perturbation 
$\mu$.}

\vskip 0.5cm
We next study the dependence on the parameters of the shell.  Fig. 3a
shows the trajectories of the lowest damping QNMs ($j=0,1,\cdots,6$) for
different $r_s$ for the case of $l=1$ scalar wave with $\mu=0.01$
based on exact numerical calculation.  We note the rich features of
the perturbed spectra.  As $r_s$ changes, the QNMs
execute complicated trajectories on the complex $\omega$ plane, with the
higher-order modes moving more rapidly as $r_s$ varies.  This
behavior is readily understood from the perturbation formula
(\ref{eq:result}), which gives

\begin{equation}
\omega_1 \sim e^{2i\omega_0 x_s}/x_s^2~~~~~\mbox{for~~~~}x_s/2M_a \gg 1.
\end{equation}

\noindent
With $\mbox{Im $\omega_0 < 0$}$,
the QNMs move away from the unperturbed positions in an
exponential
fashion as $x_s$ increases, and the higher-order modes ($- \mbox{Im }2M_a \omega_0 \gg 1$)
move with 
higher speed. 

More intriguingly there are complicated fine structures in these
trajectories.  Fig. 3b shows the fine details of the trajectory of the
$j=0$ mode.  The results obtained by direct numerical integration and
by the 1st order perturbation formula are shown.  For
larger $x_s$, the trajectory shows a spiral structure, which can be
explained from the first-order perturbation formula: 

\begin{equation}
\omega_1 = \int_{-\infty}^{x_s} H(x)V_1(x) dx +\kappa(x_s)H(x_s).
\end{equation}
The asymptotic behavior of $H(x)$ is $H(x) \sim e^{2 i \omega_0 x}$, so for 
large $x_s$,  ($'=d/dx_s$)

\begin{eqnarray}
\omega_1 ' \sim  \left[
 V_1(x_s)+\kappa'(x_s)+2i\omega_0 \kappa(x_s) \right] e^{2i\omega_0 x_s}.
\end{eqnarray} 

\noindent
The exponential factor $e^{2i\omega_0 x_s}$ gives the spirial 
structure. 

\begin{figure}
\centerline{\epsfig{file=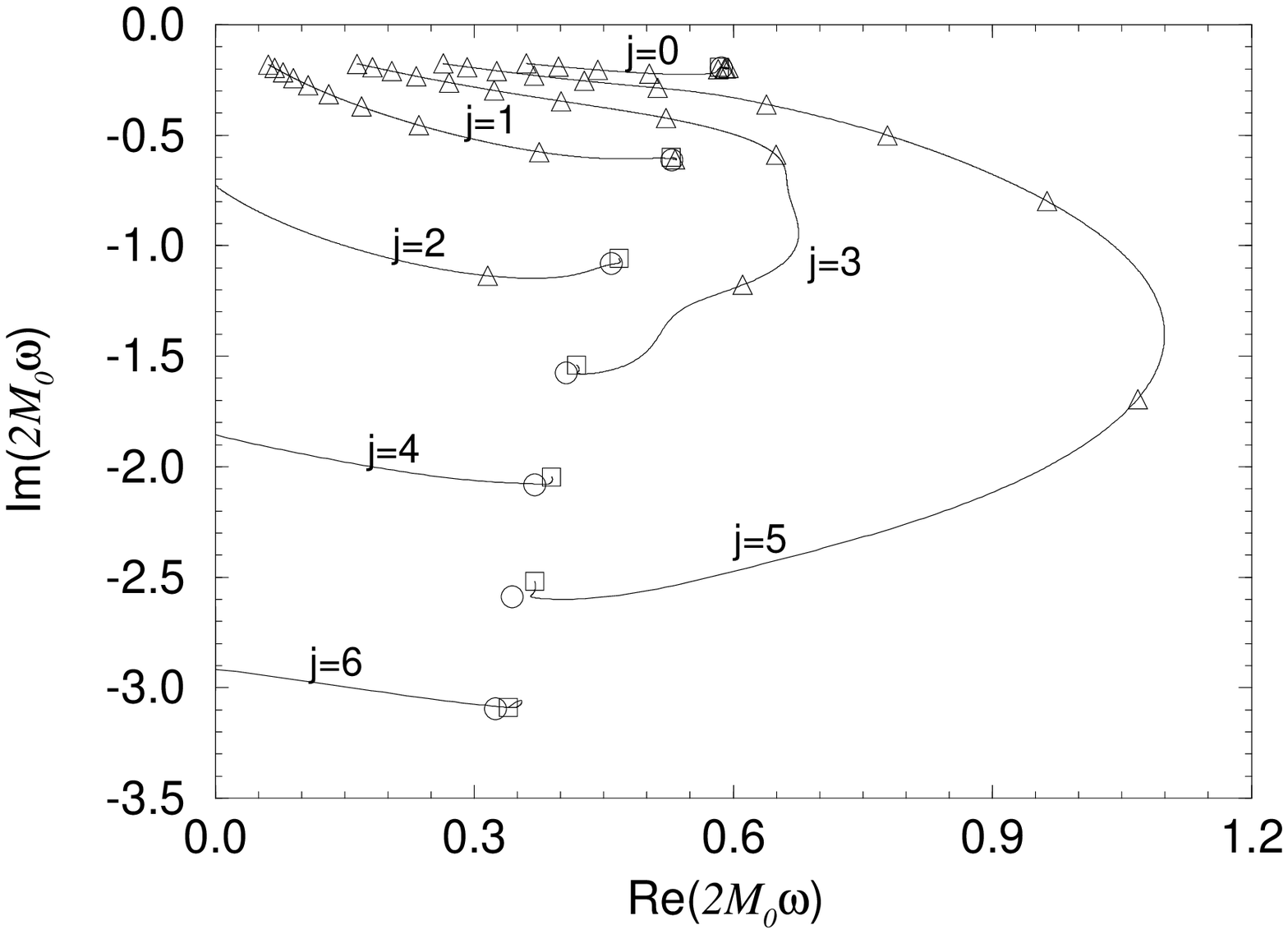,height=5.5cm}}
\end{figure}

\vskip -1.5cm
\ \ 

\begin{figure}
\centerline{\epsfig{file=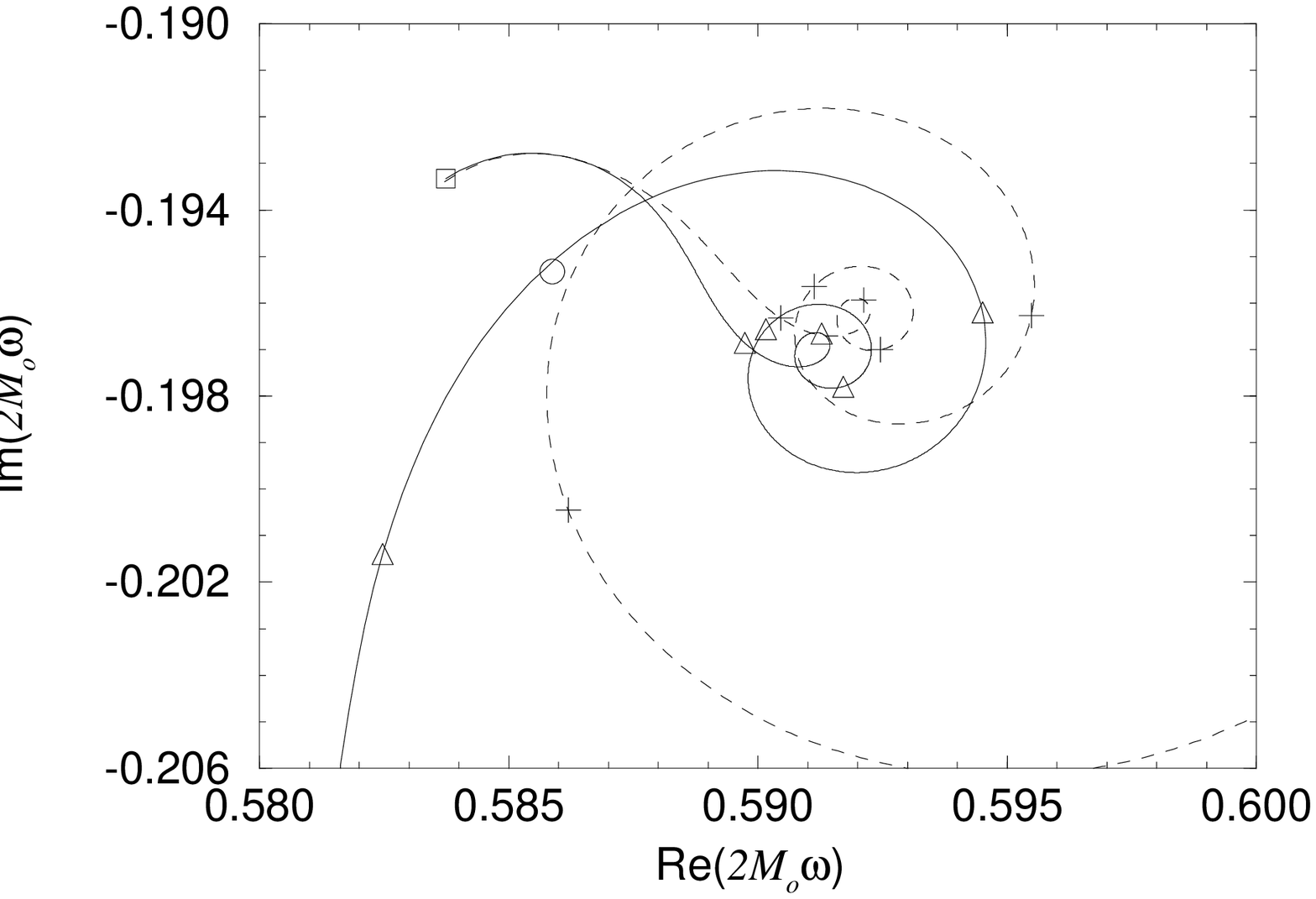,height=5.5cm}}
\end{figure}

\vskip -0.5cm
\noindent   
{\small Fig. 3a (the upper graph)~~The trajectory of the lowest QNMs of
$l=1$ scalar
waves for $\mu=0.01$ and $r_s/M_a$ varying from $2.22$ to $60$.  The
circles are QNMs of a bare Schwarzschild hole with mass $M_o$; the   
squares are the QNMs for $r_s=2.22M_a$ (The dominant energy condition
is violated when $r_s<2.22M_a$); the triangles show the positions of
QNMs at $r_s/M_a=$ $6$ to $60$ in intervals of $6$.  Fig. 3b (the lower 
graph) shows
the detail of the trajectory of the $j=0$ mode based on exact (solid 
line) and 1st order (dashed line) calculation.}

\vskip 0.5cm
{\it 6. Conclusion.}  We have developed a formulation for the
perturbation of QNMs, in close parallel to the familiar perturbation theory, 
which is directly applicable to
black holes.  With QNM gravitational wave signals from black holes to
be detected soon, and many black holes expected to be perturbed by
their astrophysical environments, e.g., accretion disks, this
formulation will be of interest to gravitational wave astronomy.

Although the QNMs of any 
system can in principle be obtained through brute force numerical
integration, perturbation formulas are often more revealing.  We note
the usefulness of perturbation theory in
conventional conservative systems,e.g., in quantum mechanics.
Moreover, the 
numerical integration of QNM spectrum is much more difficult than for
NM system.

In summary, we raise the importance of studying the QNMs of dirty
black holes, and have developed a perturbation formulation for this
purpose.  The formulation opens the way to extracting rich information
from gravitational wave signals from black hole events, and leads the
way to study of the inverse problem.  We show in a simple example that the
perturbed spectrum shows interesting features, which can be understood
with the perturbation formula.

This work is supported in part by the Hong Kong Research Grants
Council grant 452/95P, and the US NSF grant PHY 96-00507.  WMS also wants
to thank the support of the Institute of Mathematical Science of The
Chinese University of Hong Kong.
We thank C. K. Au for discussions about LPT.


\end{multicols}

\end{document}